\documentclass[a4paper,11pt]{article}
\usepackage[left=3cm,right=2cm,top=2cm,bottom=2cm]{geometry}
\usepackage[english]{babel}
\usepackage[utf8x]{inputenc}
\usepackage{amsmath,amssymb}
\usepackage{graphicx}
\usepackage{bbold}
\usepackage{color}
\usepackage{tikz}
\usepackage{multicol}
\usetikzlibrary{arrows}

\newcommand{\ov}[1]{\overline{#1}}
\newcommand{\abs}[1]{\lvert #1 \rvert}

\newtheorem{prop}{Proposition}
\newtheorem{defi}{Definition}

\title{Essential paths space on ADE $SU(3)$ graphs:\\A geometric approach}
\author{Jes\'us A. Pineda \qquad Esteban Isasi \qquad Mario I. Caicedo\\\\{\it Departamento de F\'isica, Universidad Sim\'on Bol\'ivar}\\{\it Apartado Postal 89000, Caracas 1080-A,}\\{\it Venezuela}}
\begin{document}
\maketitle

\begin{abstract}
For simply laced $SU(3)$ graphs we offer a geometric understanding of the path creation and annihilation operators for $SU(3)$ in terms of creation and annihilation of sequences of three vertices forming triangular cells or collapsed triangular cells.  We prove that the space of paths of a given length can be decomposed as a direct sum of orthogonal sub-spaces constructed by recurrent applications of the path creation operator on subspaces of essential paths of shorter length.
\end{abstract}

\section{Introduction.}

The ADE classification of $\hat{su}(N)$ WZW Conformal Field Theories (CFT) modular invariants revealed the tip of the iceberg of a rich and interesting algebraic structure, which is still being explored today. Historically the identification that led to the ADE classification was first established for $N=2$ by Cappelli, Itzykson and Zuber in \cite{Cappelli:1987xt}. Later, Terry Gannon \cite{Gannon:1992ty} provided the full list of partition functions associated to $\hat{su}(3)$, and the list of graphs generalizing the $ADE$ classification was constructed by Di Francesco and Zuber \cite{DiFrancesco1990602}  and by A. Ocneanu \cite{Ocneanu:2000}.


The formal structure that lies behind this classification has been developed over time \cite{Ocneanu:paths}, \cite{Ocneanu:2000}, \cite{CoqueGarciaTrinchero:1999},\cite{CoqueTrincheroDTE:2004}, \cite{EvansBockenhauerIII:1998}, \cite{EvansPintoSubfqctor:2003}, \cite{Coque6jsymbols:2006} and has driven progress of several important aspects of RCFTs, as, for example, the extension of the set of partition functions associated with a pair of coherent ADE graphs $\{G, A(G)\}$ beyond the modular invariant to include twisted partition functions \cite{Petkova:2000ip}. This new set of partition functions describe CFTs and statistical models lattices which admit boundary conditions on a twisted torus as well as topological defects. These models have successfully expanded into various applications \cite{Ho:2014vla}, \cite{Morin-Duchesne:2013qma}, \cite{Okuda:2014fja}. On the other hand this formal structure has developed and proven to be sufficiently interesting to become a topic of independent study connecting mathematically a huge list of, apparently disparate, topics: statistical mechanics, string theory, quantum gravity, conformal field theory, re-normalization in quantum field theory, theory of bimodules, Von Neumann algebras, sector theory, (weak) Hopf algebras, modular categories, etc.

The algebraic structures involved in the ADE classification can be summarized as follows: To every modular invariant of a 2D RCFT with chiral algebra $\hat{sl}(N)$ at level $k$ one can associate a special kind of quantum groupoid\footnote{In \cite{Ocneanu:2000} aspects of the general $su(N)$ case were discussed, and graphs families for $N = 2,3,4$ are shown. In \cite{CoquerauxSchieber:2009} the algebra of quantum symmetries is determined for the three exceptionals of $su(4)$ at levels 4, 6 and 8.} (weak Hopf algebra) $\mathcal{B}$ constructed from the combinatorial data associated to a especial pair of graphs $\{G,A(G)\}$. $G$ being an $ADE$ generalized Coxeter-Dynkin graph \cite{DiFrancesco1990602},\cite{Ocneanu:2000}, \cite{CoquereauxHammahui:2006} (when $N=2$ this means an $A,D$ or $E$ Dynkin graph) with generalized Coxeter number $\kappa=k+N$ and $A(G)$ an $A_k$ graph of the same generalized Coxeter-Dynkin system, with the same generalized Coxeter number. $A(G)$ is called the Fusion Graph of $G$. The vertices of $G$ and $A(G)$ form vector spaces, whose elements, called {\it{irreducible quantum dimensions}} can be added and multiplied much like the irreducible representations of groups. It turns out that the vector space spanned by the vertices of the graph G is a module over the graph $A(G)$, and the module action is defined by a set of structure constants $\mathcal{F}_{n,a,b}$ that encode the so called admissible triangles, diffusion graphs or also essential paths on the graph \cite{Coquereaux:2005hc}, \cite{CoquereauxGarcia:2005}, \cite{Coquereaux:2004by}. These admissible triangles form a diagrammatical representation of the well known fusion rules for the primary fields in a RCFT.

If one then takes the graded endomorphisms of admissible triangles one finds that it is possible to imbue this space $\mathcal{B}$ with the structure of an associative, coassociative, unital and counital block bialgebra. Furthermore, this block bialgebra induces a similar structure in the dual space. This second block bialgebra can be represented in a third graph called the Ocneanu graph ($Oc(G)$) which encodes the quantum symmetries of the theory. In practice, the products on $\mathcal{B}$ and its dual are naturally defined through a pair of basis which are not dual to each other\footnote{We call these basis the ``vertical'' and ``horizontal'' basis  \cite{Coquereaux:2005hc},\cite{DiFrancescoSenechal:1997}.}. As a consequence of this, the coproduct in $\mathcal{B}$ is naturally defined in a basis that is not the same as the one that gives the straightforward definition of the product on $\mathcal{B}$. This represents a computational complication that should not be underestimated, even more so as this problem is repeated for the coproduct in the dual space. The change of basis between these basis of the bialgebra is given by the set of Ocneanu cells ~\cite{Coquereaux:2005hc}. The explicit calculation of these cells can be, computationally speaking, extremely demanding, primarily because dimensionality increases rapidly once one gets past the first simplest cases. Only for a few examples \cite{
Coquereaux:2003qg}, \cite{HammahuiA2SU32008} the complete Hopf algebra has been computed.

In order to recover the connection with $2D$ RCFT we have to consider additional properties of $\mathcal{B}$. As mentioned before, $G$ is a module over $A(G)$, in the same way that $G$ is a module over $Oc(G)$ and $Oc(G)$ is a bimodule on $A(G)$. The structure constants of this bimodule define the set of Toric Matrices $W_{xy}$. The matrix $W_{00}$ is modular invariant and gives the partition function of the corresponding RCFT with chiral algebra $\hat{sl}(N)$ in terms of the Virasoro characters, and the remaining matrices $W_{x,y}$ define twisted partition functions with one $(x,0)$, $(0,y)$ or two $(x,y)$ topological defect lines, labeled by the indices $x,y$ \cite{Petkova:2000ip}. The fusion rules of the theory are given, as expected, by the Fusion Graph.

In \cite{Trinchero:2010yr} it is shown that for any simple ADE bioriented graph there exists a quantum grupoid  $\mathcal{B}$. The quantum grupoid is obtained directly from the properties of the essential paths subspace without having to calculate Ocneanu cells. The key ingredient is the decomposition of the space of paths as a direct sum of sub-spaces which are: either the subspace of essential paths of length $n$, or orthogonal subspaces constructed by recurrent applications of the corresponding creation operator $C^{\dagger}$ on subspaces of essential paths of shorter length. This decomposition and the corresponding orthogonal projectors, are sufficient to compute the quantum grupoid.

Concerning the specific properties of the subset of essential paths for the $SU(3)$ family not much is actually known, apart from a general idea of how the operators $C$ and $C^{\dagger}$ must behave \cite{CoquereauxIsasiSchieber:2010}. Their precise definitions, properties and, very important, clear and distinct interpretations for their action on paths have been lacking. The behavior of these operators differs from $SU(2)$ since $SU(3)$ graphs are oriented, and even more, each bialgebra $\mathcal{B}$ is associated to a pair of graphs $G$ and $\bar{G}$ conjugate to one another. In particular the concept of a  backtracking path, which is key in many aspects of this construction, is not as straightforward as it is in $SU(2)$ ~\cite{CaicedoIsasiPineda:2015}. 

Also the combined action of the two operators $U_{i}=C_{i}^{\dagger}C_{i}$ must define a set of Jones' projectors. We must obtain in this way a path realization of the well known Jones–Temperley–Lieb algebra modified for $SU(3)$. In \cite{CoquereauxIsasiSchieber:2010} a basic description of the operator $C_i$ and the unitary operator $C^{\dagger}_i C_i$ is given in terms the values of the triangular cells. In the same work and in \cite{Evans:2009ar} the values of this cells have been calculated for almost the complete list of graphs of $SU(3)$ family. This opens the door for the operational definition of the creation and annihilation operators in $SU(3)$, as they act on the triangles of the graph and require the values of the triangular cells.

In this work we discuss geometrically the action of the creation, cup, cap and annihilation operators and we use them to explicitly find essential paths for two examples of $SU(3)$ ADE graphs. This discussion clearly shows that the fundamental differences in the definitions and properties of the operators required to find the essential paths for $SU(2)$ and $SU(3)$ graphs are natural and inherent to the structure of the two families of graphs. We also prove that for a simply laced $ADE$ $SU(3)$ graph, the space of paths of a given length can be decomposed as a direct sum just as it happens in  the $SU(2)$ case. This represents an important first step in order to fully flesh out a path formulation for the bialgebra associated to $SU(3)$ graphs.
 

\section{Paths on $SU(3)$ graph.}
Let us now consider any generalized ADE graph $G$ of the $SU(3)$ family. Its adjacency matrix has a Perron Frobenius eigenvalue and its related unique eigenvector. This eigenvalue defines a generalized Coxeter number for $G$: $\beta=1+2\cos(2\pi/\kappa)$, and the level of $G$ is defined from the generalized Coxeter number as $k=\kappa-3$.

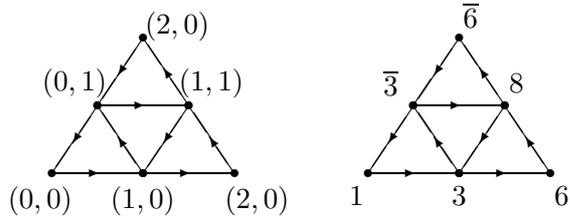
\begin{figure}
\begin{center}\setlength{\unitlength}{0.3mm}
\begin{picture}(80,80)
\put(0,0){\begin{picture}(40,40)(-10,-10)  
\put(0,0){\circle*{4}}  
\put(40,0){\circle*{4}}  
\put(20,30){\circle*{4}}  
\put(0,0){\vector(1,0){21}}  
\put(20,0){\line(1,0){20}}  
\put(40,0){\vector(-2,3){11}}  
\put(30,15){\line(-2,3){10}}  
\put(20,30){\vector(-2,-3){11}}  
\put(10,15){\line(-2,-3){10}}
\put(-5,-12){\makebox(0,0){$(0,0)$}}
\end{picture}}

\put(40,0){\begin{picture}(40,40)(-10,-10)  
\put(0,0){\circle*{4}}  
\put(40,0){\circle*{4}}  
\put(20,30){\circle*{4}}  
\put(0,0){\vector(1,0){21}}  
\put(20,0){\line(1,0){20}}  
\put(40,0){\vector(-2,3){11}}  
\put(30,15){\line(-2,3){10}}  
\put(20,30){\vector(-2,-3){11}}  
\put(10,15){\line(-2,-3){10}}
\put(0,-12){\makebox(0,0){$(1,0)$}}
\put(50,-12){\makebox(0,0){$(2,0)$}}
\end{picture}}

\put(20,30){\begin{picture}(40,40)(-10,-10)  
\put(0,0){\circle*{4}}  
\put(40,0){\circle*{4}}  
\put(20,30){\circle*{4}}  
\put(0,0){\vector(1,0){21}}  
\put(20,0){\line(1,0){20}}  
\put(40,0){\vector(-2,3){11}}  
\put(30,15){\line(-2,3){10}}  
\put(20,30){\vector(-2,-3){11}}  
\put(10,15){\line(-2,-3){10}}
\put(-10,10){\makebox(0,0){$(0,1)$}}
\put(35,35){\makebox(0,0){$(2,0)$}}
\put(50,10){\makebox(0,0){$(1,1)$}}
\end{picture}}
\end{picture}
\hspace{1.5cm}
\begin{picture}(80,80)
\put(0,0){\begin{picture}(40,40)(-10,-10)  
\put(0,0){\circle*{4}}  
\put(40,0){\circle*{4}}  
\put(20,30){\circle*{4}}  
\put(0,0){\vector(1,0){21}}  
\put(20,0){\line(1,0){20}}  
\put(40,0){\vector(-2,3){11}}  
\put(30,15){\line(-2,3){10}}  
\put(20,30){\vector(-2,-3){11}}  
\put(10,15){\line(-2,-3){10}}
\put(-5,-10){\makebox(0,0){$1$}}
\end{picture}}

\put(40,0){\begin{picture}(40,40)(-10,-10)  
\put(0,0){\circle*{4}}  
\put(40,0){\circle*{4}}  
\put(20,30){\circle*{4}}  
\put(0,0){\vector(1,0){21}}  
\put(20,0){\line(1,0){20}}  
\put(40,0){\vector(-2,3){11}}  
\put(30,15){\line(-2,3){10}}  
\put(20,30){\vector(-2,-3){11}}  
\put(10,15){\line(-2,-3){10}}
\put(0,-10){\makebox(0,0){$3$}}
\put(45,-10){\makebox(0,0){$6$}}
\end{picture}}

\put(20,30){\begin{picture}(40,40)(-10,-10)  
\put(0,0){\circle*{4}}  
\put(40,0){\circle*{4}}  
\put(20,30){\circle*{4}}  
\put(0,0){\vector(1,0){21}}  
\put(20,0){\line(1,0){20}}  
\put(40,0){\vector(-2,3){11}}  
\put(30,15){\line(-2,3){10}}  
\put(20,30){\vector(-2,-3){11}}  
\put(10,15){\line(-2,-3){10}}
\put(-10,10){\makebox(0,0){$\ov{3}$}}
\put(25,40){\makebox(0,0){$\ov{6}$}}
\put(45,10){\makebox(0,0){$8$}}
\end{picture}}
\end{picture}
\end{center}
\caption{The $SU(3)$ $\mathcal{A}$-type graph of level 2. We display two useful notations for vertices: one using so called ``triangular coordinates'' and another one using the labels of the irreducible representations of $SU(3)_{q}$).}
\label{fig:A2SU3}
\end{figure}

The normalized eigenvector associated with the biggest eigenvalue $\beta$ is called the dimension vector, which is normalized by setting to $1$ its smallest entry. If the graph is of type $\mathcal{A}$, the vertices of the graph are labelled by the so called triangular coordinates $\lambda = (\lambda_1 , \lambda_2 )$ \cite{CoquereauxHammahui:2006}, see figure (\ref{fig:A2SU3}). The quantum dimension of a vertex is given by the q-analogue of the classical formula for dimensions of $SU (3)$ irreps\footnote{In the case of an $\mathcal{A}$-type graph, these irreps are the integrable representations of the affine algebra to level k, that are the current algebra of a WZW RCFT.}, usual numbers being replaced by quantum numbers: $qdim(\lambda) = (\frac{1}{[2]_q})([\lambda_1 + 1]_q [\lambda_2 + 1]_q [\lambda_1 + \lambda_2 + 2]_q )$, where $q = exp(\frac{i\pi}{\kappa})$ is a root of unity and $[n]_q =\frac{q^{n}-q^{-n}}{q^1-q^{-1}}$. With this normalization, the quantum dimension of the unit vertex is $1$, while the 
number $\beta$ itself is the quantum dimension of the fundamental vertices, also called generators, $\sigma=(1,0)$ and $\bar{\sigma}=(0,1)$ \cite{CoquereauxHammahui:2006,CoquereauxIsasiSchieber:2010}.

\begin{defi}
\label{defi-path}
An elementary path is a sequence of vertices  $\eta=v_{0}v_{1}\dots v_{i-1}v_{i}v_{i+1}\dots v_{n}$ connected by arrows which may belong to either one graph or its {\it{conjugate}}. An elementary path can also be defined as a series of consecutive edges on any of the graphs.
\end{defi}

 Formally, an elementary path is built through a succession of actions of either generator, $\sigma$ and $\ov{\sigma}$ \cite{CoquereauxIsasiSchieber:2010}. Since $SU(3)$ graphs are bidimensional and oriented one finds that there are ambiguities in the definition of length. However, it is possible to introduce a definition of length that consists in counting the edges of each type in the path, in \cite{CaicedoIsasiPineda:2015} it is shown that all definitions of length involving the counting of generators are interchangeable. Here we classify elementary paths with an ordered pair of integers $(\alpha,\beta)$ that counts the amount of edges generated by $\sigma$ and $\ov{\sigma}$. 

\begin{defi} The length of a path is $n=(\alpha+\beta)$, where $(\alpha,\beta)$ are integers that give the number of edges generated by $\sigma$ and $\ov{\sigma}$ respectively. As usual, this is equivalent to the total number of vertices minus 1.
\end{defi}

This labelling scheme implies that paths of the same length can have differing values for both indices and can therefore be of different type. Also, if the graph $G$ is of type $A$, of level $l$ and if $\alpha+\beta\leq l$ then $(\alpha,\beta)$ coincide with the triangular coordinates $(\lambda_{1},\lambda_{2})$ of the $A$ type graph.

\begin{defi}
The inner product vector space of paths $\mathcal{P}$ is defined by saying that elementary paths provide a orthonormal basis of this space. This space is naturally graded in the following way:
\begin{equation}
\mathcal{P}=\bigoplus_{n}\mathcal{P}^{(n)}=\bigoplus_{n}\bigoplus_{\alpha+\beta=n}\mathcal{P}_{(\alpha,\beta)}.
\end{equation}
\end{defi}

A particular subspace $\mathcal{E}$ of the space of paths is constituted by the so called {\it{essential paths}} \cite{Ocneanu:paths}. In $SU(2)$, essential paths can be roughly defined as those without backtrackings. Here the notion of a backtracking segment of a path is straightforward: any section that goes back and forth between a vertex and one of its neighbors is a backtracking segment. The set of all essential paths is defined as the kernel of all annihilation operators. We now move forward to recreate these definitions for $SU(3)$ graphs.

\subsection{Creation and annihilation operators}

The ADE $SU(3)$ family naturally lives on a 2D lattice, the vertices of the graphs are subsets of the Weyl alcove of $su(3)$ at level $k$. The elements of this family are therefore characterized by graphs based on triangular cells, this property means that there are two geometrically natural ways of producing a round trip starting and ending on the same vertex. The first one implies going to one of the vertex nearest neighbours and back (exactly as in the $SU(2)$ case). The second one requires a round trip through two additional vertices that, alongside the original one, form a closed triangular cell. 

A backtracking sequence of vertices of a path over a $SU(3)$ graph, is either a sequence of three vertices $v_{i-1}v_{i}v_{i+1}$ that form a triangle or a back and forth sequence where $v_{i-1}=v_{i+1}$. This implies that every triangular backtracking sequence can be associated to a triangular cell $T_{i-1,i,i+1}$ that assigns a complex number to each elementary triangle of the graph. This assignment is sometimes called “a self-connection”, “a connection on the system of triangular cells”, or simply “a cell system” and the $T_{i-1,i,i+1}$ cells have already been  defined and calculated in \cite{Evans:2009ar, CoquereauxIsasiSchieber:2010, CaicedoIsasiPineda:2015}. This cell system is determined via a set of nontrivial equations that represent the Kuperberg identities between the $SU(3)$ intertwiners. 

We now introduce the creation, annihilation, cup and cap operators:
\begin{defi}
Given a path $\eta=v_{0}v_{1}\dots v_{i-1}v_{i}v_{i+1}\dots v_{n}$:
{\footnotesize
\begin{multicols}{2}
\begin{equation}\label{eq:defaniq}
C_{i}(\eta)= \frac{T_{i-1,i,i+1}}{\sqrt{[i-1][i+1]}}v_{0}v_{1}\dots v_{i-1}v_{i+1}\dots v_{n},
\end{equation}
\begin{equation}\label{eq:defcrea}
C^{\dag}_{i}(\eta)= \sum_{b\,n.n.\,v_i}\frac{\overline{T_{i-1,b,i}}}{\sqrt{[i-1][i]}}v_{0}v_{1}\dots v_{i-1}v_{b}v_{i}\dots v_{n}.
\end{equation}

\begin{equation}\label{eq:defcup}
\cup_{i}(\eta)= \frac{T_{i-1,i,i+1}}{\sqrt{[i-1][i+1]}}\delta_{i-1,i+1}\;v_{0}v_{1}\dots v_{i-1}v_{i+2}\dots v_{n},
\end{equation}
\begin{equation}\label{eq:defcap}
\cap_{i}(\eta)= \sum_{v_b \,n.n.\,v_i}\frac{\overline{T_{i-1,b,i-1}}}{\sqrt{[i-1][i-1]}}v_{0}v_{1}\dots v_{i-1}v_{b}v_{i-1}v_{i}\dots v_{n}.
\end{equation}
\end{multicols}}
\end{defi}

Where the prefactor that was $\sqrt{\mu_{v_{i+1}}/\mu_{v_{i-1}}}$ for a single $SU(2)$ elementary backtrack  is now replaced by an $SU(3)$ elementary triangle $\frac{T_{i-1,i,i+1}}{\sqrt{[i-1][i+1]}}$ which ensures that these operators satisfy the Temperley-Lieb algebra. The values of these coefficients for elementary triangles  have been calculated in \cite{Evans:2009ar, CoquereauxIsasiSchieber:2010}. 

\begin{defi}
A collapsed triangular sequence  is  a back and forth sequence of three vertices $v_{i-1},v_{i},v_{i+1}$ such that  the first an last vertices ($v_{i-1}$ and $v_{i+1}$) are the same.
\end{defi}

\begin{prop}
Given a collapsed triangular sequence $v_{i-1},v_{i},v_{i+1}$, its associated cell $T_{i-1, i, i-1}$ is given by 
\begin{equation}
T_{i-1, i, i-1}=\sqrt{[i-1][i]}.
\end{equation}
\end{prop}
This results from calculating the relation $(C_{i}C^{\dag}_{i})^{2}$ which, using the properties of the triangular cells (i.e. the Kuperberg relations) yields the result $\mathbb{1}+\cup_{i}\cap_{i}$ from which one can read the values of the collapsed cells associated with the cup and cap operators.

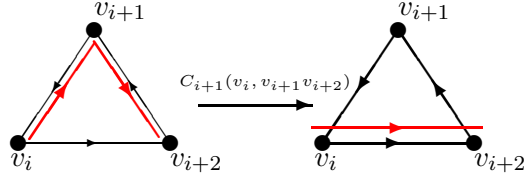
\begin{figure}[hhh]
\centering
\setlength{\unitlength}{0.5mm}
\begin{picture}(100,50)(0,-10)
\put(0,0){
\begin{picture}(40,40)  
\put(0,0){\circle*{4}}  
\put(40,0){\circle*{4}}  
\put(20,30){\circle*{4}}  
\put(0,0){\vector(1,0){21}}  
\put(20,0){\line(1,0){20}}  
\put(40,0){\vector(-2,3){11}}  
\put(30,15){\line(-2,3){10}}  
\put(20,30){\vector(-2,-3){11}}  
\put(10,15){\line(-2,-3){10}}
\put(-2,-6){$v_i$}
\put(20,34){$v_{i+1}$}
\put(40,-6){$v_{i+2}$}
\thicklines
\color{red}{
\put(3,1){\vector(2,3){10}}
\put(3,1){\line(2,3){17}}
\put(20,27){\vector(2,-3){10}}  
\put(20,27){\line(2,-3){17}}
}
\end{picture}
}
\put(45,15){{\tiny $C_{i+1}(v_i,v_{i+1}v_{i+2})$}}
\thicklines
\put(50,10){\vector(1,0){30}}
\put(80,0){
\begin{picture}(40,40)  
\put(0,0){\circle*{4}}  
\put(40,0){\circle*{4}}  
\put(20,30){\circle*{4}}  
\put(0,0){\vector(1,0){21}}  
\put(20,0){\line(1,0){20}}  
\put(40,0){\vector(-2,3){11}}  
\put(30,15){\line(-2,3){10}}  
\put(20,30){\vector(-2,-3){11}}  
\put(10,15){\line(-2,-3){10}}
\put(-2,-6){$v_i$}
\put(20,34){$v_{i+1}$}
\put(40,-6){$v_{i+2}$}
\thicklines
\color{red}{
\put(-3,4){\vector(1,0){25}}
\put(22,4){\line(1,0){20}}
}
\end{picture}
}
\end{picture}
\label{aniquilacion}
\caption{The action of annihilation operator on a section of path that passes through three vertices forming an $SU(3)$ backtrack. The operator removes one of the vertices and connects the remaining two through one shorter path. The red arrow is notes the way the section of the path is constructed.}
\end{figure}

\begin{figure}[hhh]
\centering
\setlength{\unitlength}{0.5mm}
\begin{picture}(200,100)(0,-35)
\put(0,0){
\begin{picture}(40,40)  
\put(0,0){\circle*{4}}  
\put(40,0){\circle*{4}}  
\put(20,30){\circle*{4}}  
\put(20,-30){\circle*{4}}

\put(0,0){\vector(1,0){21}}  
\put(20,0){\line(1,0){20}}  

\put(40,0){\vector(-2,3){11}}  
\put(30,15){\line(-2,3){10}}  

\put(40,0){\vector(-2,-3){11}}  
\put(30,-15){\line(-2,-3){10}}  

\put(20,30){\vector(-2,-3){11}}  
\put(10,15){\line(-2,-3){10}}

\put(20,-30){\vector(-2,3){11}}  
\put(10,-15){\line(-2,3){10}}

 \put(-2,-6){$v_i$}
 \put(20,34){$v_{b1}$}
 \put(40,-6){$v_{i+1}$}
 \put(20,-35){$v_{b2}$}
\thicklines
\color{red}{
\put(-3,4){\vector(1,0){25}}
\put(22,4){\line(1,0){20}}
}
\end{picture}
}
\put(48,15){{\tiny $C^{\dagger}_{i+1}(v_i,v_{i+1})$}}
\thicklines
\put(50,10){\vector(1,0){30}}
\put(80,0){
\begin{picture}(40,40)  
\put(0,0){\circle*{4}}  
\put(40,0){\circle*{4}}  
\put(20,30){\circle*{4}}
\put(20,-30){\circle*{4}}

\put(0,0){\vector(1,0){21}}  
\put(20,0){\line(1,0){20}}  

\put(40,0){\vector(-2,3){11}}  
\put(30,15){\line(-2,3){10}}  

\put(40,0){\vector(-2,-3){11}}  
\put(30,-15){\line(-2,-3){10}}

\put(20,30){\vector(-2,-3){11}}  
\put(10,15){\line(-2,-3){10}}

\put(20,-30){\vector(-2,3){11}}  
\put(10,-15){\line(-2,3){10}}

\put(-2,-6){$v_i$}
\put(20,34){$v_{b1}$}
\put(40,-6){$v_{i+1}$}
\put(20,-35){$v_{b2}$}
\thicklines
\color{red}{
\put(3,1){\vector(2,3){10}}
\put(3,1){\line(2,3){17}}
\put(20,27){\vector(2,-3){10}}  
\put(20,27){\line(2,-3){17}}
}
\end{picture}
}
\put(140,0){\Large{$+$}}
\put(160,0){
\begin{picture}(40,40)  
\put(0,0){\circle*{4}}  
\put(40,0){\circle*{4}}  
\put(20,30){\circle*{4}}
\put(20,-30){\circle*{4}}

\put(0,0){\vector(1,0){21}}  
\put(20,0){\line(1,0){20}}  

\put(40,0){\vector(-2,3){11}}  
\put(30,15){\line(-2,3){10}}  

\put(40,0){\vector(-2,-3){11}}  
\put(30,-15){\line(-2,-3){10}}

\put(20,30){\vector(-2,-3){11}}  
\put(10,15){\line(-2,-3){10}}

\put(20,-30){\vector(-2,3){11}}  
\put(10,-15){\line(-2,3){10}}

\put(-2,-6){$v_i$}
\put(20,34){$v_{b1}$}
\put(40,-6){$v_{i+1}$}
\put(20,-35){$v_{b2}$}
\thicklines
\color{red}{
\put(3,-1){\vector(2,-3){10}}
\put(3,-1){\line(2,-3){17}}
\put(20,-27){\vector(2,3){10}}  
\put(20,-27){\line(2,3){17}}
}
\end{picture}
}
\end{picture}
\caption{The action of creation operator on a section of path that passes through two consecutive vertices  of a path with neighbouring vertices $v_{b1}$ and $v_{b2}$. The operator ads the two combinations passing through the neighbouring vertices connecting the two original vertices with a longer path. The red arrow is notes the way the section of the path is constructed.}
\label{creation}
\end{figure}
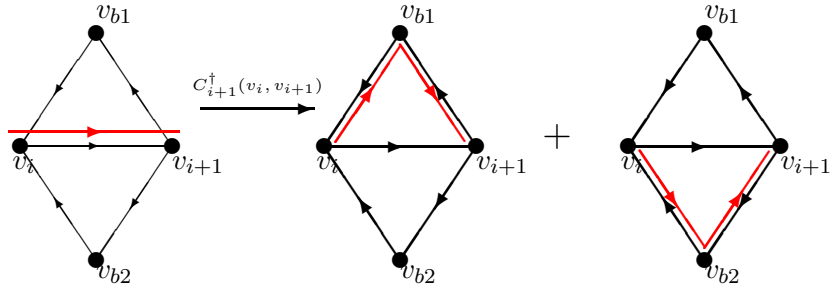

The creation and annihilation operators defined in \ref{eq:defaniq} and \ref{eq:defcrea} add or remove ``open'' triangular sequences of vertices in such a way that the vertex that is added or removed is the middle one of the triangle constructed with the vertices $i-1,i,i+1$. If such a sequence does not exist or if the operator acts on a path of length $n<2$, the annihilation operator $C_{i}$ gives zero result. For the creation operator the sum is over all vertices $v_{b}$ that are common neighbors of $i-1$ and $i$. Here we see that the meaning of a backtracking sequence for $SU(2)$ changes to a triangular sequence in $SU(3)$.  In figures \ref{aniquilacion} and \ref{creation} one can see a geometrical representation of the action of both operators on a sequence of vertices of a path.

The cup and cap operators in equations \ref{eq:defcup} and \ref{eq:defcap} create or annihilate back-and-forth sequences that are not clearly triangular but are in direct analogy to the back-and forth sequences in $SU(2)$ \cite{Trinchero:2010yr}. Whenever the input path lacks one such back-and-forth sequences the action of the cup operator on it yields zero. The sum in the cap operator adds all possible back-and-forth sequences to the nearest neighbors of the $i-1$-th vertex. These operators make use of the so called collapsed triangular cells, that were introduced above. The creation and annihilation operators, and the cup and cap operators provide a complete set of operators that fulfill the Kuperberg relations for rank 2 tangles. See \cite{CaicedoIsasiPineda:2015} for details.  

We expect that this set of four operators will provide a set of essential paths whose properties should be either equivalent or very close to those found found for $SU(2)$, i.e., the operators should fulfill the following properties:
\begin{enumerate}
\item The object $U_i=C_i^{\dagger}C_i$ should fulfill the $SU(3)$ version of the Temperley--Lieb algebra.
 \item The kernel of the annihilation and cup operators $C_i$, $\cup_{i}$ must provide the full set of essential paths; this set matching the one defined from the module action of $A(G)$ on $G$.
\end{enumerate}

We now check that the definitions for the creation and annihilation operators satisfy the $SU(3)$ Temperley-Lieb algebra \cite{Evans:2009ud}, that is to say, that the set of operators $U_{i}=C_{i}\,C_{i}^{\dagger}$, provides a realization of the following relations:
\begin{align}
U_{i}^{2}&=\beta U_{i},\label{H1-xx}\\
U_{i}U_{j}&=U_{j}U_{i},\quad \abs{i-j}>1,\label{H2-xx}\\
F_{i}=U_{i}U_{i+1}U_{i}-U_{i}&=U_{i+1}U_{i}U_{i+1}-U_{i+1},\label{H3-xx}\\
(U_{i}-U_{i+2}U_{i+1}U_{i}+U_{i+1})&(U_{i+1}U_{i+2}U_{i+1}-U_{i+1})=0.\label{H4-xx}
\end{align}
This $U_i$ operator has already been defined in \cite{CoquereauxIsasiSchieber:2010} and is related to the rhombii that can be built combining triangles in the graph.

The action of the $U_{i}$ operator is to initially remove vertex $i$ and then add a new vertex in such a way that two paths are created: the original path $\eta$ and a second path in which the $i$-th vertex is replaced by another one which, together with those in the original sequence $v_{i-1}v_{i}v_{i+1}$, form a rhombus with the edge connecting the $i-1$ and $i+1$ vertices as a diagonal, i.e. a pair of elementary triangles that share the aforementioned edge. 

The proof of equations \ref{H1-xx} and \ref{H2-xx} follow from the definition of $U_{i}$. Relation (\ref{H3-xx}) is interesting since it allows us to see that our definitions in fact subsume closed triangular sequences. Whenever the path is not a closed triangle (i.e. that is one containing a sequence $v_{i-1}v_{i}v_{i+1}v_{i-1}$) the relation is trivially satisfied, which is to say the rhombii produced by either sequence of $U$ operators is exactly the same. However, if the path contains a closed triangular sequence  we are naturally led to a definition for the $F_{i}$ operator previously studied in \cite{Evans:2009ud,CoquereauxIsasiSchieber:2010}:
\begin{align}
F_{i}(\eta)&=\sum_{b',b''}\frac{T_{i-1,i,i+1}\ov{T_{i-1,b'',b'}}}{[i-1][i-1]}\;v_{0}v_{1}\dots v_{i-1}v_{b''}v_{b'}v_{i-1}\dots v_{n}\nonumber\\
&=\sum_{b',b''}\frac{T_{i,i+1,i-1}\ov{T_{b',b'',i-1}}}{[i-1][i-1]}\;v_{0}v_{1}\dots v_{i-1}v_{b''}v_{b'}v_{i-1}\dots v_{n}=F_{i+1}(\eta).
\end{align}
In this way we find that all straightforward reinterpretations of backtracking paths in $SU(3)$ (i.e. a path going from a vertex to a neighbor and back, and one taking a triangular round trip from a vertex to itself) are all accounted for, flowing naturally from our operators. In \cite{CaicedoIsasiPineda:2015} it is shown that back-and-forth sequences and closed triangles are interchangeable as backtrackings in $SU(3)$.

The final relation (\ref{H4-xx}), which defines the Temperley-Lieb algebra for $SU(3)$, has already been shown to be satisfied whenever the following simpler relation is fulfilled (cf. \cite{Evans:2009ud} Lemma 4.1): $F_{i}F_{i+1}F_{i}=\beta^{2}F_{i}$. Once more, the proof can be obtained by direct calculation.

\begin{defi}
\label{defi-essential-1}
A path $\eta$ is essential if:
 \begin{equation}
 C_{i}(\xi)=0,\quad\cup_{i}(\eta)=0,\qquad\mbox{both for all}\; i<n.
 \end{equation}
 and its length is $n=\alpha+\beta$ where $(\alpha,\beta)$  is a vertex of the $A$-type graph that shares the same generalized Coxeter number of the original graph $G$.
\end{defi}

Essential paths of length $(\alpha ,0)$ and $(0,\beta)$ were already calculated by Evans and Pugh \cite{EvansPughCalabiYau:2012}. Our geometric approach which led us to find a definition of the collapsed triangular cells allowed a full understanding of backtracking paths, which in turn ended up in the above definition which encompasses paths of any length.

The module action $\mathcal{A}(\mathcal{G})\times \mathcal{G}\rightarrow\mathcal{G}$ defines the admissible triangles $(anb)$ which is associated to existence of an essential path going from vertex $a$ to vertex $b$ and having length given by the label $n=(\alpha,\beta)$ of the $\mathcal{A}(\mathcal{G})$ graph. For any admissible triangle there exists at least one linear combination (up to gauge fixing and given values for the set of triangular cells) of elementary paths that begin at vertex $a$ and end at vertex $b$ all of which have length $n$ and which gets mapped to zero under the action of $C$.

\subsection{Examples}
\label{Examples}

\paragraph{$G=A_2$}

$A$ type graphs exhibit self-fusion (which implies that the multiplication table coincides with the module action), an explicit calculation of the fusion product table is straightforward and is displayed below for the $A_{2}$ graph.

\begin{table}[h]
$$
\begin{array}{c|cccccc}
\nearrow & 1 & 3 & 6 &  \ov{3} &  \ov{6} & 8\\\hline
 1 & 1 & 3 & 6 &  \ov{3} &  \ov{6} & 8\\
 3 & 3 &  \ov{3}+6 & 8 & 1+8 &  \ov{3} &  \ov{6}+3\\
 6 & 6 & 8 &  \ov{6} & 3 & 1 &  \ov{3}\\
  \ov{3} &  \ov{3} & 1+8 & 3 &  \ov{6}+3 & 8 & 6+ \ov{3}\\
  \ov{6} &  \ov{6} &  \ov{3} & 1 & 8 & 6 & 3\\
 8 & 8 &  \ov{6}+3 &  \ov{3} & 6+ \ov{3} & 3 & 1+8
\end{array}
$$
\caption{$A_2$ multiplication table.}
\label{A2multiplication}
\end{table}

Table \ref{A2multiplication} allows for the construction of the space of admissible triangles, wherein the triplet $(n,k,m)$ will be admissible if the m-eth representation appears in the table as the decomposition in irreducible sums of the tensor product of $n\otimes k$. Additionally, the table encodes the essential paths of all admissible sizes over the graph, as was shown in \cite{Trinchero:2010yr}. The full expressions of the essential paths in terms of elementary paths and their linear combinations are obtained through the action of the annihilation and cup operators (see Table \ref{esspathtab1}).

\begin{table}[h]
\begin{center}
\begin{tabular}{c|c|c}
Path size & Essential paths& Generators\\\hline
$(0,0)$ & $(1)$, $(3)$, $(\ov{3})$, $(6)$, $(\ov{6})$, $8$ & $\sigma^0$ \\
$(1,0)$ & $(13)$, $(3\ov{3})$, $(36)$, $(\ov{3}1)$, $(\ov{3}8)$, $(68)$, $(\ov{6}\ov{3})$, $(8\ov{6})$, $(83)$ & $\sigma^1$ \\
$(0,1)$ & $(1\ov{3})$, $(\ov{3}3)$, $(\ov{3}\ov{6})$, $(31)$, $(38)$, $(\ov{6}8)$, $(63)$, $(86)$, $(8\ov{3})$  & $\ov{\sigma}^1$ \\
$(2,0)$ & \parbox[c][1.5cm]{7cm}{$(68\ov{6})$, $(\ov{6}\ov{3}1)$, $(136)$, $(368)-\sqrt{[2]}(3\ov{3}8)$, $(\ov{3}83)-\sqrt{[2]}(\ov{3}13)$,  $(8\ov{6}\ov{3})-\sqrt{[2]}(83\ov{3})$} & $\sigma^2$ \\
$(0,2)$ &  \parbox[c][1.5cm]{7cm}{$(\ov{6}86)$, $(631)$, $(1\ov{3}\ov{6})$, $(\ov{3}\ov{6}8)-\sqrt{[2]}(\ov{3}38)$, $(38\ov{3})-\sqrt{[2]}(31\ov{3})$, $(863)-\sqrt{[2]}(8\ov{3}3)$} & $\ov{\sigma}^2$ \\
$(1,1)$ &  \parbox[c]{6.5cm}{$(138)$, $(1\ov{3}8)$, $(831)$, $(8\ov{3}1)$, $(3\ov{3}\ov{6})$, $(38\ov{6})$, $(\ov{6}\ov{3}3)$, $(\ov{6}83)$, $(\ov{3}36)$, $(\ov{3}86)$, $(63\ov{3})$, $(68\ov{3})$, $(313)-\sqrt{\frac{[1]}{[8]}}(383)+(3\ov{3}3)-\sqrt{\frac{[\ov{3}]}{[6]}}(363)$, $(\ov{3}1\ov{3})-\sqrt{\frac{[1]}{[8]}}(\ov{3}8\ov{3})+(\ov{3}3\ov{3})-\sqrt{\frac{[3]}{[\ov{6}]}}(\ov{3}\ov{6}\ov{3})$, $(8\ov{6}8)-\sqrt{\frac{[\ov{6}]}{[3]}}(838)+(868)-\sqrt{\frac{[6]}{[\ov{3}]}}(8\ov{3}8)$ }& $\sigma\ov{\sigma}, \ov{\sigma}\sigma$
\end{tabular}
\caption{List of essential paths ordered in terms of their length and the number of generators that produce them.}
\label{esspathtab1}
\end{center}
\end{table}

Since there is a maximum length for essential paths given by the $A$ type graph with the same generalized Coxeter number as $G$ we do not list as essential paths of greater length. This can also be understood by noticing that the triple action of either generator $\sigma$ or $\ov{\sigma}$ gives a reducible representation containing $1$ and $8$ (e.g. $(3\cdot 3)\cdot 3=(\ov{3}+6)\cdot 3=\ov{3}\cdot 3+6\cdot 3=1+8+8$, where symmetry ensures a similar result for all corners of the graph), in geometrical terms this means that a path with three edges following the arrows renders one of three cases: first, a closed triangular path, and two instances of paths that can be reduced to paths of length $(1,1)$. However, in $SU(3)$, one can act with either $\sigma$ or $\ov{\sigma}$, which implies that paths longer than the essential path of maximal length, and belonging in the kernel of the annihilation operator, can be produced. These are explicitly the paths $(138\ov{6})$, $(68\ov{3}1)$,  $(\ov{6}\ov{3}36)$ and the equivalent three generated by conjugation. As a consequence of this, the preceding list of essential paths is complete only if one takes as essential only paths given by the set of admissible triangles and of length lesser than or equal to that given by the Coxeter number of the graph\footnote{Since this is the case we will not  be concerned by these longer paths.}.

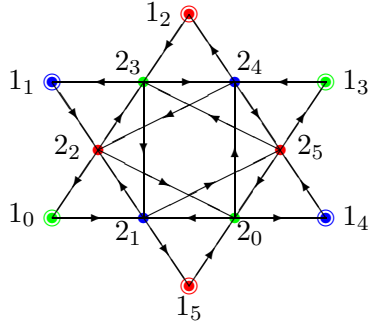
\begin{figure}[hhh] 
\begin{center} 
\unitlength 0.2mm 
\begin{picture}(200,200)

\put(0,20){\begin{picture}(180,180)

\put(0,45){\begin{picture}(60,45) 
\put(0,0){\color{green} \circle*{7}} 
\put(0,0){\color{green} \circle{10}} 
\put(60,0){\color{blue} \circle*{7}} 
\put(30,45){\color{red} \circle*{7}} 
\put(0,0){\vector(1,0){32.5}} 
\put(30,0){\line(1,0){30}} 
\put(60,0){\vector(-2,3){16.5}} 
\put(45,22.5){\line(-2,3){15}} 
\put(30,45){\vector(-2,-3){16.5}} 
\put(15,22.5){\line(-2,-3){15}} 
\end{picture}} 
 
\put(120,45){\begin{picture}(60,45) 
\put(0,0){\color{green} \circle*{7}} 
\put(60,0){\color{blue} \circle{10}} 
\put(60,0){\color{blue} \circle*{7}} 
\put(30,45){\color{red} \circle*{7}} 
\put(0,0){\vector(1,0){32.5}} 
\put(30,0){\line(1,0){30}} 
\put(60,0){\vector(-2,3){16.5}} 
\put(45,22.5){\line(-2,3){15}} 
\put(30,45){\vector(-2,-3){16.5}} 
\put(15,22.5){\line(-2,-3){15}} 
\end{picture}} 
 
\put(60,135){\begin{picture}(60,45) 
\put(0,0){\color{green} \circle*{7}} 
\put(60,0){\color{blue} \circle*{7}} 
\put(30,45){\color{red} \circle*{7}} 
\put(30,45){\color{red} \circle{10}} 
\put(0,0){\vector(1,0){32.5}} 
\put(30,0){\line(1,0){30}} 
\put(60,0){\vector(-2,3){16.5}} 
\put(45,22.5){\line(-2,3){15}} 
\put(30,45){\vector(-2,-3){16.5}} 
\put(15,22.5){\line(-2,-3){15}} 
\end{picture}}

\put(0,90){\begin{picture}(60,45) 
\put(0,45){\color{blue} \circle*{7}} 
\put(0,45){\color{blue} \circle{10}} 
\put(60,45){\vector(-1,0){32.5}} 
\put(30,45){\line(-1,0){30}} 
\put(30,0){\vector(2,3){16.5}} 
\put(45,22.5){\line(2,3){15}} 
\put(0,45){\vector(2,-3){16.5}} 
\put(15,22.5){\line(2,-3){15}} 
\end{picture}} 
 
\put(120,90){\begin{picture}(60,45) 
\put(60,45){\color{green} \circle*{7}} 
\put(60,45){\color{green} \circle{10}} 
\put(60,45){\vector(-1,0){32.5}} 
\put(30,45){\line(-1,0){30}} 
\put(30,0){\vector(2,3){16.5}} 
\put(45,22.5){\line(2,3){15}} 
\put(0,45){\vector(2,-3){16.5}} 
\put(15,22.5){\line(2,-3){15}} 
\end{picture}} 
 
\put(60,0){\begin{picture}(60,45) 
\put(30,0){\color{red} \circle*{7}} 
\put(30,0){\color{red} \circle{10}} 
\put(60,45){\vector(-1,0){32.5}} 
\put(30,45){\line(-1,0){30}} 
\put(30,0){\vector(2,3){16.5}} 
\put(45,22.5){\line(2,3){15}} 
\put(0,45){\vector(2,-3){16.5}} 
\put(15,22.5){\line(2,-3){15}} 
\end{picture}} 
 
\put(60,135){\vector(0,-1){47.5}} 
\put(60,90){\line(0,-1){45}} 
\put(60,45){\vector(2,1){47.2}} 
\put(105,67.5){\line(2,1){45}} 
\put(150,90){\vector(-2,1){47.2}} 
\put(105,112.5){\line(-2,1){45}} 
 
\put(120,45){\vector(0,1){47.5}} 
\put(120,90){\line(0,1){45}} 
\put(120,135){\vector(-2,-1){47.2}} 
\put(75,112.5){\line(-2,-1){45}} 
\put(30,90){\vector(2,-1){47.2}} 
\put(75,67.5){\line(2,-1){45}}

\put(-20,47){\makebox(0,0){$1_0$}} 
\put(200,45){\makebox(0,0){$1_4$}} 
\put(-20,135){\makebox(0,0){$1_1$}} 
\put(200,135){\makebox(0,0){$1_3$}} 
\put(90,-15){\makebox(0,0){$1_5$}} 
\put(70,180){\makebox(0,0){$1_2$}} 

\put(50,35){\makebox(0,0){$2_1$}}  
\put(130,35){\makebox(0,0){$2_0$}} 
\put(50,148){\makebox(0,0){$2_3$}} 
\put(130,148){\makebox(0,0){$2_4$}} 
\put(10,92){\makebox(0,0){$2_2$}} 
\put(170,92){\makebox(0,0){$2_5$}} 

\end{picture}} 
\end{picture} 
\caption{The $SU(3)$ $\mathcal{E}_{5}$ graph.} 
\label{E5SU3} 
\end{center} 
\end{figure} 

\paragraph{$G=E_5$}
This is a level 5 graph and its fusion algebra is given by the action of an $A_{5}$ graph. For brevity, we avoid the fusion product and the module product tables (which can be found in \cite{IsasiSchieber:2007, CoquereauxHammahui:2006}). A similar procedure as  before yields the essential paths of which we provide a list up to length $n=2$ and a few examples of length $n=0+3$.

\begin{table}
\begin{center}
\begin{tabular}{c|c|c}
Path size & Essential paths & Generators\\\hline
$(0,0)$ & $(1_{i})$, $(2_{i})$ & $\sigma^{0}$\\
$(1,0)$ & $(1_{i}2_{i+1})$, $(2_{i}2_{i+1})$, $(2_{i},2_{i+4})$, $(2_{i}1_{i+4})$ & $\sigma^{1}$\\
$(0,1)$ & $(1_{i}2_{i+2})$, $(2_{i}2_{i+2})$, $(2_{i},2_{i+5})$, $(2_{i}1_{i+5})$ & $\ov{\sigma}^{1}$\\
$(1,1)$ & \parbox[c][2.8cm]{7.5cm}{$(1_{i}2_{i+1}2_{i})$, $(1_{i}2_{i+2}2_{i})$, $(1_{i}2_{i+2}2_{i+3})$, $(1_{i}2_{i+1}2_{i+3})$, $(2_{i}2_{i+5}2_{i})+(2_{i}2_{i+2}2_{i})-2\sqrt{\frac{[2_{i+2}]+[2_{i+5}]}{2[1_{i+5}]}}(2_{i}1_{i+5}2_{i})$, $(2_{i}2_{i+5}2_{i+3})$, $(2_{i}2_{i+1}2_{i+3})$} & $\sigma\ov{\sigma}$,$\ov{\sigma}\sigma$\\
$(2,0)$ & \parbox[c][2.8cm]{7.5cm}{$(1_{i}2_{i+1}2_{i+5})$, $(1_{i}2_{i+1}1_{i+5})$, $(2_{i}2_{i+4}1_{i+2})$, $(2_{i}2_{i+4}2_{i+2})-\frac{\nu_{0}}{\mu}(2_{i}2_{i+1}2_{i+2})$, $(2_{i}1_{i+4}2_{i+5})-\frac{\mu}{\tau}(2_{i}2_{i+1}2_{i+5})$, $(2_{i}1_{i+4}2_{i+5})-\frac{\mu}{\tau}(2_{i}2_{i+4}2_{i+5})$} & $\sigma^{2}$\\
$(0,3)$ & \parbox[c][2cm]{7cm}{$(1_{i}2_{i+2}2_{i+4}1_{i+3})$, $(1_{i}2_{i+2}2_{i+4}2_{i})-\frac{\nu_{0}}{\mu}(1_{i}2_{i+2}2_{i+1}2_{i})$ ($i$ even),  $(1_{i}2_{i+2}2_{i+4}2_{i})+\frac{\nu_{1}}{\mu}(1_{i}2_{i+2}2_{i+1}2_{i})$ ($i$ odd)} & $\ov{\sigma}^{3}$ 
\end{tabular}\caption{Partial list of essential paths for $E_5$, sorted by length and number of generators.}\label{tabesspathe5}
\end{center}
\end{table}

In table \ref{tabesspathe5} we have listed families of paths: due to the symmetries of the graph one can find sets of paths that share the same structure and are related by rotations and reflections, therefore the index $i$ in the table run from $0$ to $5$ and all operations are $\pmod{6}$. Using conjugation over $E_{5}$\footnote{This amounts to symmetry with respect to the axis joining vertices $1_{0}$ and $1_{3}$ leading to the relations $\ov{1}_{0}=1_{0}$, $\ov{1}_{5}=1_{1}$, $\ov{1}_{4}=1_{2}$, $\ov{1}_{3}=1_{3}$, $\ov{2}_{0}=2_{3}$, $\ov{2}_{1}=2_{2}$ and $\ov{2}_{5}=2_{4}$} one can find the list of the essential paths complementary to those listed here (i.e. those of length $(0,2)$ and $(3,0)$). 

For clarity, we show some explicit calculations of the action of the annihilation operator on some paths of the $E_5$ graph. The path $(1_{3}2_{4}1_{2})$ should be essential:
\begin{equation}
C_{1}(1_{3}2_{4}1_{2})= \frac{T_{1_{3}2_{4}1_{2}}}{\sqrt{[1_{3}][1_{2}]}}\;(1_{3}1_{2})=0.
\end{equation}
The result is easy to reach since the resulting path is disjoint. For another, slightly longer path ($n=(2,1)$), $(1_{3}2_{4}2_{3}1_{1})$:
\begin{equation}
C_{1}(1_{3}2_{4}2_{3}1_{1})= \frac{T_{1_{3}2_{4}2_{3}}}{\sqrt{[1_{3}][2_{3}]}}\;(1_{3}2_{3}1_{1})=0,\quad C_{2}(1_{3}2_{4}2_{3}1_{1})= \frac{T_{2_{4}2_{3}1_{1}}}{\sqrt{[2_{4}][1_{1}]}}\;(1_{3}2_{4}1_{1})=0.
\end{equation}
Both paths then, belong to the kernel of the annihilation operators and we can thus claim that, for a given path length, some linear combination comprising the above paths conforms the basis of the space of essential paths. For a path that we can easily tell is not essential $(1_{3}2_{4}2_{3}2_{2})$ we get:
\begin{equation}
C_{2}(1_{3}2_{4}2_{3}2_{2})= \frac{T_{2_{4}2_{3}2_{2}}}{\sqrt{[2_{4}][2_{2}]}}\;(1_{3}2_{4}2_{2}).
\end{equation}
Which is to be expected given that the triplet $2_{4}2_{3}2_{2}$ forms a triangle in the graph and thus results in a nonessential path.

\section{Decomposition of the space of paths}

We will now show that it is possible to decompose the space of paths of a given length in a manner analogous to the $SU(2)$ case (cf.~\cite{Trinchero:2010yr} equation [5.7]).

\begin{prop}
The space of paths of length $n$ that connect the vertices $a$ and $b$, $\mathcal{P}^{(n)}_{ab}$, can be decomposed as follows: 
\begin{multline}
\mathcal{P}^{(n)}_{ab}= \mathcal{E}^{(n)}_{ab}\bigoplus_{i\leq n-2} (C^{\dag}+\cap)_{i}(\mathcal{E}^{(n-1)}_{ab})\bigoplus_{i_{1}<i_{2}\leq n-2}(C^{\dag}+\cap)_{i_{2}}(C^{\dag}+\cap)_{i_{1}}(\mathcal{E}^{(n-2)}_{ab})\bigoplus\;\dots\\ \dots\bigoplus_{i_{1}<i_{2}\dots < i_{[n/2]}\leq n-2} (C^{\dag}+\cap)_{i_{[n/2]}}(C^{\dag}+\cap)_{i_{[n/2]-1}}\dots (C^{\dag}+\cap)_{i_{1}}(\mathcal{E}^{(1|0)}_{ab}).
\end{multline}
where $\mathcal{E}^{(n)}_{ab}$ is the space of essential paths of length $n$ connecting vertices $a$ and $b$ and $(C^{\dag}+\cap)_{i}(\mathcal{E}^{(n-j)}_{ab})=C^{\dag}_{i}(\mathcal{E}^{(n-j)}_{ab})\bigoplus\cap_{i}(\mathcal{E}^{(n-j)}_{ab})$ is the space spanned by the paths obtained by the action of the creation and cap operator on the elements of the space of essential paths of length $n-j$.
\end{prop}

{\bf Proof:} In order to prove that such a decomposition of the space of paths of length $n$ connecting two vertices exists we will proceed through induction. We start by defining a family of operators that create backtracking segments acting on elementary paths (of which our creation and cap operators are examples). If $\eta=v_{0}v_{1}\dots v_{i-1}v_{i}v_{i+1}\dots v_{n}$ is an elementary path, then  $c_{i}^{\dag}(\eta)=\sum_{b}\beta_{b}v_{0}v_{1}\dots v_{i-1}v_{b}v_{i}v_{i+1}\dots v_{n}$. This definition allows us to select a particular elementary path with a backtracking segment in the result of the action of the creation or cap operator on a given path $\eta$ (or a selection thereof) using appropriate $\beta$ coefficients. Clearly, the spaces $C^{\dag}(\mathcal{E}_{n})\bigcup\cap(\mathcal{E}_{n})$ and $c^{\dag}(\mathcal{E}_{n})$ are equal once we allow for all possible values of the $\beta$ coefficients. Since the only relevant fact for us at this point is to have a way to create nonessential elementary 
paths of length $n+1$, that is, elements of the base of paths of length $n+1$ which are, by definition, orthonormal to the elementary path $\eta$ (of length $n$) from which we obtained them.  

Let $\eta_{ab}^{(n)}$ be a path of length $n$ connecting vertices $a$ and $b$. Since elementary paths provide an orthonormal basis of the space of paths we can write any path of a given length as 
\begin{equation}\label{decompelempath}
\eta_{ab}^{(n)}=\sum_{i}\alpha_{i}e_{ab}^{(n)\; i},
\end{equation}
where $\{e_{ab}^{(n)}\}$ is the basis of the space of elementary paths of length $n$ connecting vertex $a$ to vertex $b$. For elementary paths one can define the concatenation product $\bullet$ such that for two elementary paths $\eta=v_{0}v_{1}\dots v_{i}$, $\eta'=v'_{j}v'_{1}\dots v'_{p}$ one gets $\eta\bullet\eta'=\delta_{v_{i},v'_{j}}\;v_{0}v_{1}\dots v_{i}v'_{1}\dots v'_{p}$. This allows us to split an elementary path $e_{ab}^{(n)}=e_{ac}^{(n-l)}\bullet e_{cb}^{(l)}$ in such a way that $(C|\cup)_{k}(e_{cb}^{(l)})=0$ for all $k>n-l$ and $(C|\cup)_{k}(e_{ac}^{(n-l)})\neq 0$ for at least one $k\leq n-l-2$. Here the notation $(C|\cup)_{k}(\eta)$ means that one is to act on $\eta$ with the annihilation or cup operators according to the type of backtracking section of the path.

Given the decomposition of elementary paths in equation \ref{decompelempath} we can then rewrite the $e_{ac}^{(n-l)}$ subpath, which is elementary but nonessential as one resulting from the action of the creation or cap operator defined above with suitable values for the $\beta$ coefficients, thus $e_{ac}^{(n-l)}=c_{k}^{\dag}(e_{ac}^{(n-l-1)})$. This path of length $n-l-1$ is now such that $c_{k'}(e_{ac}^{(n-l-1)})\neq 0$ for some $k'<n-l-3$ since we have removed the leftmost ``backtracking'' from the sequence. We can now replicate this procedure on this new elementary path by cutting off the rightmost essential subpath using the concatenation product: $\eta_{ac}^{(n-l-1)}=e_{af}^{(n-l-1-m)}\bullet e_{fc}^{(m)}$, now with $(C|\cup)_{k}(e_{fc}^{(l)})=0$ for all $k>n-l-m-3$ and $(C|\cup)_{k}(e_{ac}^{(n-l)})\neq 0$ for at least one $k\leq n-l-m-2$. This implies that our initial elementary subpath can be written as 
\begin{equation}
e_{ac}^{(n-l)}= c_{k}^{\dag}\left(c_{k'}^{\dag}\left(e_{af}^{(n-l-m-1)}\right)\bullet e_{fc}^{(m)}\right),
\end{equation}
which in turn makes our original elementary path $e_{ab}^{(n)}$:
\begin{equation}
e_{ab}^{(n)}= \left(c_{k}^{\dag}\left(c_{k'}^{\dag}\left(e_{af}^{(n-l-m-1)}\right)\bullet e_{fc}^{(m)}\right)\right)\bullet e_{cb}^{(l)}.
\end{equation}
We can repeat this procedure until we reach a rightmost subpath that is the result of the application of the creation or cap operator on a path of size $0$ (i.e. a vertex) or $1$. This results an expression for an elementary path of length $n$ connecting $a$ to $b$ (and thus for a general path of the same length connecting the same vertices) as a sequence of ordered applications of the creation or cap operator on an initial path in the kernel of the annihilation or cup operator which is then concatenated with a series of essential subpaths, explicitly:
\begin{equation}
e_{ab}^{(n)}= \left(c_{k}^{\dag}\left(c_{k'}^{\dag}\left(c_{k''}^{\dag}\dots\left(c_{k^{(s)}}^{\dag}e_{ap_{1}}^{(r)}\right)\bullet e_{p_{1}p_{2}}^{(s)}\right)\bullet e_{p_{2}p_{3}}^{(s')}\right)\dots\right)\bullet e_{cb}^{(l)}.
\end{equation}
Where all the $e_{ij}$ are subpaths in the kernel of $(C|\cup)_{i}$ and in particular the innermost in the above equation can be of length $0$ or $1$ at the very smallest. As we can see this sequence of applications of the $c^{\dag}$ operators is unique since the ordering in our procedure results in only one such sequence for a given path and since the original path remains one of length $n$ we can conclude that any path of length $n$ connecting two vertices can be written as above. Since spaces $C^{\dag}(\mathcal{E}_{n})\bigcup\cap(\mathcal{E}_{n})$ and $c^{\dag}(\mathcal{E}_{n})$ are equal the above procedure leads us to a natural decomposition for the space of paths of a given length connecting two vertices. $\blacksquare$

\section{Discussion}
Using triangular sequences of vertices as an analogue for a backtracking path in $SU(3)$ we have provided a geometrical approach for backtracking paths that is both straightforward and natural given the building blocks of the family of $SU(3)$ graphs. This geometric understanding of a backtracking path in $SU(3)$ leads us to grasp the meaning of the path creation, annihilation, cup and cap operators in terms of the creation and annihilation of triangular or back-and-forth sequences of vertices. 

In addition we have shown, through the explicit calculation of some examples for both $A$ type and $E$ type graphs, that our definitions of the annihilation and cup operators and our geometrical interpretations, yield not just the desired results for essential and nonessential paths but also set up interesting questions that could allow for a generalization of this work for the remaining members of the $SU(3)$ family, that is to say, the $D$ series (multiply connected) and the conjugate graphs (as these graphs can be understood as constructed with collapsed triangles).

Our main result is the decomposition of the space of paths: we have found that the space of paths of a given length can be decomposed in a way similar to that shown previously for $SU(2)$, which is to say that one can obtain a path of a given length by taking a path in the kernel of the annihilation and cup operators and then acting upon it with an ordered sequence of creation and cap operators until one obtains the desired path. What the above decomposition does not provide, is an explicit way of writing a given path in terms of a linear combination of elementary paths of equal length or of shorter length on which one has acted with a specific ordering of creation and cap operators. In future works we will propose a first idea for an algorithm that takes a path and explicitly deconstructs it in elements of the subspaces described above. A similar algorithm has already been fully explored for $SU(2)$ in \cite{Trinchero:2010yr}. 

\bibliographystyle{unsrt}
\bibliography{bibliography-paths-su3} 

\begin{thebibliography}{10}

\bibitem{Cappelli:1987xt}
Andrea Cappelli, C.~Itzykson, and J.B. Zuber.
\newblock {The ADE Classification of Minimal and A1(1) Conformal Invariant
  Theories}.
\newblock {\em Commun.Math.Phys.}, 113:1, 1987.

\bibitem{Gannon:1992ty}
Terry Gannon.
\newblock {The Classification of affine SU(3) modular invariant partition
  functions}.
\newblock {\em Commun.Math.Phys.}, 161:233--264, 1994.

\bibitem{DiFrancesco1990602}
P.~Di Francesco and J.-B. Zuber.
\newblock Su(n) lattice integrable models associated with graphs.
\newblock {\em Nuclear Physics B}, 338(3):602 -- 646, 1990.

\bibitem{Ocneanu:2000}
Adrian Ocneanu.
\newblock The classification of subgroups of quantum {${\rm SU}(N)$}.
\newblock In {\em Quantum symmetries in theoretical physics and mathematics
  (Bariloche, 2000)}, volume 294 of {\em Contemp. Math.}, pages 133--159. Amer.
  Math. Soc., Providence, RI, 2002.

\bibitem{Ocneanu:paths}
Adrian Ocneanu.
\newblock Paths on coxeter diagrams: fron platonic solids and singularities to
  minimal models and subfactors.
\newblock In {\em Lectures on Operator Theory}, volume~33 of {\em Fields
  Institute Monographs}, pages 245--323. American Mathematical Society, 1999.

\bibitem{CoqueGarciaTrinchero:1999}
R.~Coquereaux, A.~O. Garcia, and R.~Trinchero.
\newblock Hopf stars, twisted hopf stars and scalar products on quantum spaces.
\newblock {\em J. Geom. Phys.}, 36:22--59, 2000.

\bibitem{CoqueTrincheroDTE:2004}
Robert Coquereaux and Roberto Trinchero.
\newblock On quantum symmetries of ade graphs.
\newblock {\em Adv. Theor. Math. Phys.}, 8:189--216, 2004.

\bibitem{EvansBockenhauerIII:1998}
Jens Bockenhauer and David~E. Evans.
\newblock Modular invariants, graphs and alpha-induction for nets of
  subfactors. iii.
\newblock {\em Commun. Math. Phys.}, 205:183--228, 1999.

\bibitem{EvansPintoSubfqctor:2003}
D.~E. Evans and P.~R. Pinto.
\newblock Subfactor realisation of modular invariants.
\newblock {\em Commun. Math. Phys.}, 237:309--363, 2003.

\bibitem{Coque6jsymbols:2006}
Robert Coquereaux.
\newblock Racah - wigner quantum 6j symbols, ocneanu cells for a(n) diagrams,
  and quantum groupoids.
\newblock 2005.

\bibitem{Petkova:2000ip}
V.B. Petkova and J.B. Zuber.
\newblock {Generalized twisted partition functions}.
\newblock {\em Phys.Lett.}, B504:157--164, 2001.

\bibitem{Ho:2014vla}
Wen~Wei Ho, Lukasz Cincio, Heidar Moradi, Davide Gaiotto, and Guifre Vidal.
\newblock {Edge-entanglement spectrum correspondence in a non-chiral
  topological phase, and Kramers-Wannier duality}.
\newblock 2014.

\bibitem{Morin-Duchesne:2013qma}
Alexi Morin-Duchesne, Paul~A. Pearce, and Jorgen Rasmussen.
\newblock {Modular invariant partition function of critical dense polymers}.
\newblock {\em Nucl.Phys.}, B874:312--357, 2013.

\bibitem{Okuda:2014fja}
Takuya Okuda.
\newblock {Line operators in supersymmetric gauge theories and the 2d-4d
  relation}.
\newblock 2014.

\bibitem{CoquerauxSchieber:2009}
R.~{Coquereaux} and G.~{Schieber}.
\newblock {Quantum Symmetries for Exceptional SU(4) Modular Invariants
  Associated with Conformal Embeddings}.
\newblock {\em SIGMA}, 5:44, April 2009.

\bibitem{CoquereauxHammahui:2006}
R.~{Coquereaux}, D.~{Hammaoui}, G.~{Schieber}, and E.~H. {Tahri}.
\newblock {Comments about quantum symmetries of SU(3) graphs}.
\newblock {\em Journal of Geometry and Physics}, 57:269--292, December 2006.

\bibitem{Coquereaux:2005hc}
Robert Coquereaux.
\newblock {Racah-Wigner quantum 6j symbols, Ocneanu cells for A(N) diagrams,
  and quantum groupoids}.
\newblock {\em J.Geom.Phys.}, 57:387--434, 2007.

\bibitem{CoquereauxGarcia:2005}
R.~{Coquereaux} and A.~O. {Garc{\'{\i}}a}.
\newblock {On Bialgebras Associated with Paths and Essential Paths on Ade
  Graphs}.
\newblock {\em International Journal of Geometric Methods in Modern Physics},
  2:441--466, June 2005.

\bibitem{Coquereaux:2004by}
Robert Coquereaux and Roberto Trinchero.
\newblock {On quantum symmetries of ADE graphs}.
\newblock {\em Adv.Theor.Math.Phys.}, 8:189--216, 2004.

\bibitem{DiFrancescoSenechal:1997}
P.~Di~Francesco, P.~Mathieu, and D.~Senechal.
\newblock {\em Conformal field theory}.
\newblock Springer, 1999.

\bibitem{Coquereaux:2003qg}
Robert Coquereaux.
\newblock {The A(2) Ocneanu quantum groupoid}.
\newblock 2003.

\bibitem{HammahuiA2SU32008}
D.~{Hammaoui}.
\newblock {The smallest Ocneanu quantumgrupoid of $SU(3)$ type}.
\newblock {\em AJSE}, 33:99, December 2008.

\bibitem{Trinchero:2010yr}
R.~Trinchero.
\newblock {Paths on graphs and associated quantum groupoids*}.
\newblock {\em Revista de la Unión Matemática Argentina}, 51:147--170, 2010.

\bibitem{CoquereauxIsasiSchieber:2010}
R.~{Coquereaux}, E.~{Isasi}, and G.~{Schieber}.
\newblock {Notes on TQFT Wire Models and Coherence Equations for SU(3)
  Triangular Cells}.
\newblock {\em SIGMA}, 6:99, December 2010.

\bibitem{CaicedoIsasiPineda:2015}
M.~Caicedo, E.~Isasi, and J.~A. Pineda.
\newblock {To be published}.
\newblock 2015.

\bibitem{Evans:2009ar}
D.~E. {Evans} and M.~{Pugh}.
\newblock {Ocneanu Cells and Boltzmann Weights for the SU(3) ADE Graphs}.
\newblock {\em Munster J. Math. 2 (2009), 95-142}, 2:95--142, June 2009.

\bibitem{Evans:2009ud}
David~E. Evans and Mathew Pugh.
\newblock {SU(3)-Goodman-de la Harpe-Jones subfactors and the realisation of
  SU(3) modular invariants}.
\newblock {\em Rev.Math.Phys.}, 21:877--928, 2009.

\bibitem{EvansPughCalabiYau:2012}
David~E. Evans and Mathew Pugh.
\newblock {The Nakayama Automorphism of the Almost Calabi-Yau Algebras
  Associated to SU(3) Modular Invariants}.
\newblock {\em Communications in Mathematical Physics}, 312(1):179--222, 2012.

\bibitem{IsasiSchieber:2007}
E.~{Isasi} and G.~{Schieber}.
\newblock {From modular invariants to graphs: the modular splitting method}.
\newblock {\em Journal of Physics A Mathematical General}, 40:6513--6537, June
  2007.

\end{thebibliography}

\end{document}